\documentstyle[preprint,prl,aps]{revtex} 

\begin{document}
\preprint{version of November 22, 1994}
\draft
\date{\today}
\title{Exact results for a one dimensional $t-J$ model with impurity}

\author{P.-A.\ Bares}
\address{
Institut Laue-Langevin, B.P.~156x, F-38042 Grenoble Cedex, France}

\maketitle%
\begin{abstract}%

We propose a simple idea
to construct 1D integrable models with
impurities.
We illustrate the strategy for a supersymmetric $t-J$ Hamiltonian
in considerable detail.
The impurity comprises the local deformation of the hopping
and exchange integrals as well as a
three-body charge-current interactions on
neighboring sites.
We explore the thermodynamic properties of the system
at low and high temperatures, and obtain results that are
beyond boundary conformal-field theory.

\end{abstract}


Recently there has been new developments in the
theory of impurities in Fermi and Luttinger liquids. This renewal of interest
in the quantum impurity problems has been stimulated in part
by the search for
non-Fermi-liquid fixed-points beyond one space dimension and its plausible
relevance to the theory of heavy fermions and high temperature
superconductors.
In this context,
Affleck, Ludwig and collaborators have used the (multi-channel) Kondo problem
as a laboratory to test, develop and extend significantly the
techniques of conformal field theory in the presence of boundaries
\cite{Cardy,AffleckCollabor}.
These powerful methods allow to obtain single-particle as well as
thermodynamic properties near the
critical point.
Using simpler bosonization and renormalization group techniques,
Kane and Fisher have
shown that a potential scatterer
embeded in a Luttinger liquid is driven to a strong-coupling fixed point
by the electron-electron repulsive interactions \cite{KaneFisher}.
Emery and Kivelson were able to map
the two channel Kondo problem to a resonant-level Hamiltonian which
could be solved exactly at a particular point\cite{EmeryKivelson}.
Fabrizio and Gogolin have generalized the Emery-Kivelson solution to
the isotropic four channel and together with Nozi\`eres to the
anisotropic multi-channel Kondo problem
\cite{FabrizioGogolin,FabGogoNozieres}.
Combining bosonization
and conformal field theory results, Sengupta and Georges have proposed
a solution for a two-channel two-impurity Kondo Hamiltonian
\cite{SenguptaGeorges}.
Significant progress in a different direction
has also been achieved by Ghoshal and
Zamolodchikov. These authors were able to solve
exactly integrable field theories with boundaries \cite{GhoshalZamolodchikov}.
Interesting applications of these results have been performed by
Fendley et al. and Tsvelick \cite{FendleyLS}.

Despite this important progress, the problem of an (few)
impurity(ies)
(potential, magnetic or boths) embeded in a strongly correlated system
is not well understood. There is a notable exception: a spin $S>1/2$
impurity in a spin $1/2$ Heisenberg chain solved many years ago by Andrei and
Johannesson \cite{AndreiJohannesson} and generalized to arbitrary spins by
Lee and Schlottmann, and Schlottmann \cite{LeeSchlottmann}. Very recently,
it was argued in reference \cite{Sorensenetal} that these models correspond
to non-generic fixed points.
In the present letter, we want to explore a new idea that allows
us to construct exactly solvable models with a (few) defect(s). Short
range interactions between the impurities can also be incorporated.
There is a prize to pay, however, namely some terms in the Hamiltonian
are difficult to justify on physical grounds. Their presence is
required by formal reasons (integrability) and should be
generically irrelevant to the low-energy properties of the system.
The idea is simple and can be explained as follows.
Consider an integrable model
that can be formulated within the framework
of the Quantum Inverse Scattering Method (QISM)\cite{QISM}.
In the QISM, the Hamiltonian of the model is
written as the logarithmic derivative of an homogeneous
transfer matrix at a special
value of the spectral parameter (an infinite number of conserved kinematic
quantities can be generated by performing successive derivatives of
the logarithm of the transfer matrix).
In general,
however, the transfer matrix
may depend on a set of (unequal) parameters. In the statictical-mechanics
language this corresponds to an inhomogeneous vertex model.
Define the Hamiltonian
as the logarithmic derivative of the corresponding inhomogeneous
transfer matrix.
Provided we restrict to an (few) inhomogeneity(ies), we have a model with
a (few) defect(s) that is exactly integrable by construction.
We have performed preliminary calculations in a variety of models:
spinless fermions, spin chains and strongly correlated systems.
A general discussion will appear elsewhere \cite{Bares}.

In this work, we want to illustrate the idea by focusing
on one particular model that is a prototype for
strongly correlated electrons, i.e., the $t-J$ model.
In the absence of impurity, the $t-J$ model is known to be solvable at the
supersymmetric point (away from the integrable point, much is also known
through numerical studies, see ref. \cite{OgataCollabor}) and various
properties of the model have been investigated in the recent past
\cite{Schlottmann,BaresCollabor,KawakamiYang}.
To construct a $t-J$ model with impurity, we follow
the strategy proposed above and
arrive at the Hamiltonian:
$H=H_0+H_I$,
where $H_0=\sum_{n=1}^{L} H_{n n-1}$ denotes the usual $t-J$ Hamiltonian,
\begin{equation}
H_{n n+1}= {\cal P} \{ - \sum_{\sigma} (c_{n\sigma}^{\dagger}c_{n+1\sigma}
+h.c.) + J({\bf S}_{n+1}\cdot {\bf S}_{n}- {{\widehat n}_n
{\widehat n}_{n+1}\over 4})+{\widehat n}_n+{\widehat n}_{n+1}
\} {\cal P}\;\;,
\end{equation}
where ${\cal P}$ projects out the doubly occupied sites,
${\widehat n}_n=\sum_{\sigma}c_{n\sigma}^{\dagger}c_{n\sigma}$, $J=2$ and
periodic boundary conditions have been assumed.
The impurity contribution can be written as
\begin{equation}
H_I= g (H_{m-1 m+1}- H_{m m-1}-H_{m m+1} )
+ i \sqrt{g(1-g)} \lbrack H_{m-1 m+1},H_{m m-1}\rbrack\;\;,
\label{hamimp}
\end{equation}
where the coupling constant $0\leq g\leq 1$ (this restriction follows from
the hermiticity of $H$) and $m$ denotes the center
of the defect that involves, besides $m$, its two adjacent sites $m\pm 1$.
The first term on the r.h.s. of (\ref{hamimp}) enhances antiferromagnetic
correlations between $m\pm1$ sites while ferromagnetic alignment is
favored between $m$ and its two adjacent sites.
The second term corresponds to
a three-body interaction,
\begin{eqnarray}
&{\cal P}& \{ ( 1 -n_m)
\sum_{\sigma}(j_{m+1 m-1}^{\sigma\sigma}+h.c.)
- (1-n_{m-1})\sum_{\sigma}(j_{m+1 m}^{\sigma\sigma}+h.c.)
-(1-n_{m+1})\sum_{\sigma}(j_{m m-1}^{\sigma\sigma}+h.c.)\nonumber\\
&+& \sum_{\sigma\sigma'}(c_{m\sigma}^{\dagger}c_{m\sigma'}
j_{m+1 m-1}^{\sigma'\sigma}+h.c.)
-\sum_{\sigma\sigma'}(c_{m-1\sigma}^{\dagger}c_{m-1\sigma'}
j_{m+1 m}^{\sigma'\sigma}+h.c.)
-\sum_{\sigma\sigma'}(c_{m+1\sigma}^{\dagger}c_{m+1\sigma'}
j_{m m-1}^{\sigma'\sigma}+h.c.)\nonumber\\
&-&\sum_{\sigma\sigma'} (c_{m\sigma}^{\dagger}c_{m\sigma'}
J_{m-1 m+1}^{\sigma\sigma'}+h.c)\} {\cal P} \;\;,
\label{threebody}
\end{eqnarray}
with
$j_{a b}^{\sigma\sigma'}=i c_{a\sigma}^{\dagger}c_{b\sigma'}$
and
$J_{a b}^{\sigma\sigma'}=i \sum_{\lambda}
(c_{a\sigma'}^{\dagger}c_{a\lambda}) (c_{b\lambda}^{\dagger}c_{b\sigma})$.
In (\ref{threebody}), we have three types of terms: (i) a local charge current
flowing between the sites $m,\;m\pm 1$ provided one of them
is empty; (ii) assisted spin-flip currents;
(iii) pure assisted spin-currents.
At $g=1$, the three body term is absent and the model behaves as a chain
of $L-1$ sites with no impurity present.

The eigenvalue problem associated to $H$
is solved by the graded version of the QISM
(see ref. \cite{EsslerKorepin}). For convenience,
we choose the Sutherland representation (see discussion in
\cite{BaresCollabor}).
The formal solution is casted into the form of
the so-called Bethe ansatz equations,
\begin{equation}
L p_{\alpha}(\lambda_j^{\alpha})+\delta_{\alpha}(\lambda_j^{\alpha})
=2\pi I_j^{\alpha}-\sum_{\beta=c,s}\sum_{k=1}^{N_{\beta}}
\phi_{\alpha\beta}^{0}(\lambda_j^{\alpha}-\lambda_k^{\beta})\;\;,
\label{BA}
\end{equation}
where $p_{\alpha}(\lambda_j^{\alpha})$, for $\alpha=c,s$,
denotes the bare momentum with $p_s(x) =2 \tan^{-1}(2 x)$ and $p_c(x)=0$,
$\phi_{\alpha\beta}^0 (x)$ is the bare phase shift for the scattering of the
$\alpha$ with the $\beta$ pseudoparticles ($\phi_{ss}^0 (x)=-2 \tan^{-1}(x)$,
$\phi_{sc}^0 (x)=\phi_{cs}^0 (x)=2 \tan^{-1}(2x)$,$\phi_{cc}^0 (x)=0 $)
while $\delta_{\alpha}(x)$ denotes the bare phase shift for the scattering
of the $\alpha$ pseudoparticle with the impurity, i.e.,
$\delta_{s}(x)= 2 [\tan^{-1}(2[x-\theta ]) -\tan^{-1}(2x)] $ and
$\delta_{c}(x)=0$. The parameter $\theta$ is defined by
$\theta=(g/(1-g))^{1/2}$.
As usual the quantum numbers $I_j^{\alpha}$ are integers or
half-odd integers depending on the number of sites $L$, the number of particles
$N$ and the magnetization $M$ in the system.
The ground-state (of finite momentum ) and part of the
low-energy spectrum are parametrized by
real roots in equation (\ref{BA}) (see ref.\cite{BaresCollabor}).

Since we are interested in thermodynamic properties,
we shall directly work at
finite temperature with the thermodynamic potential per unit length
$\Omega (\mu, H, T)$,
where $\mu$ denotes the chemical potential, $H$ the external magnetic field and
$T$ the temperature.
At finite $T$, we have to consider
all possible solutions of the set of equations (\ref{BA}).
It turns out that the
rapidities associated to the s-pseudoparticles $\lambda_j^s$
cluster into strings
in much the same way
as they do in the Heisenberg chain \cite{Bethe} while the
``charge ''rapidities $\lambda_j^c$ remain real. Following Yang and Yang
\cite{YangYang}, we write down
the thermodynamic potential
and minimize with respect to the distribution
of roots of the set of equations (\ref{BA}).
After some algebra (see for example \cite{TsvelickWiegmann}),
the thermodynamic potential per unit length, i.e., the pressure, is found,
up to an irrelevant constant,
\begin{equation}
\Omega= -\mu - h +{1\over \beta}\sum_{n=1}^{\infty}\int dx
\left[ a_n(x)+{1\over L}\left( a_n(x-\theta)-a_n(x)\right)\right]
\log \left[ 1- f(\epsilon_{n,s}(x))\right]\;,
\label{potential}
\end{equation}
where $h=\mu_B H$, $\mu_B$ is the Bohr magneton, $\beta = (k_B T)^{-1}$,
$a_n(x)= (2\pi)^{-1}dp_n(x)/dx$ with
$p_n(x)=2 \tan^{-1}(2x/n)$.
$\Omega$ naturally separates into a bulk contribution
of $O(1)$ and an impurity contribution of $O(1/L)$.
The function $f(\epsilon_{sn})=(1+e^{\beta \epsilon_{sn}})^{-1}$
coincides with the Fermi distribution function in the
low-temperature limit while it differs from the
former at finite $T$ where the quasi-energies
$\epsilon_{ns}(x)$ are $T$-dependent.
The latter
are determined from the integral equations
\begin{mathletters}
\begin{eqnarray}
\beta\epsilon_{sn}(x)
&=& s\star \log\left( f(\epsilon_{sn-1})f(\epsilon_{sn+1})\right)(x)\nonumber\\
&+& \delta_{n,1} \left[ s\star \log\left(1- f(\epsilon_{c})\right)(x)
-2\pi\beta J s(x) \right]\;\;,
\label{thermoa}\\
\beta\epsilon_{c}(x)
&=& \beta \mu - \beta h +\sum_{n} a_n\star
\log \left(1- f(\epsilon_{sn})\right)(x)\;\;,
\label{thermob}
\end{eqnarray}
\end{mathletters}%
where $s(x)=1/ 2\cosh(\pi x)$, $\star$ denotes a convolution and
the boundary conditions,
$\lim_{n\rightarrow\infty}(\epsilon_{ns}/n)=h$, have been imposed.
Notice that (\ref{thermoa}) and (\ref{thermob}) do not depend on
the impurity because the contribution of the latter to the
free energy vanishes in the
variation with respect to the densities of roots.

In general, the set of coupled equations (\ref{thermoa}) and
(\ref{thermob}) are difficult
to solve analytically. In the following, we solve these
equations in the low- and high-$T$
limits. We use techniques similar to those developped by
Takahashi\cite{Takahashi}.
We first discuss the high temperature region ($\beta J\gg 1$),
i.e., the weak coupling limit.
To lowest order in $\beta J$, the pseudo-Fermi distribution functions
$f(\epsilon_c)$ and $f(\epsilon_{sn})$ are
constants, and (\ref{thermoa}) and (\ref{thermob}) simplify to
$f(\epsilon_{sn})=F_n^{-2}$ and $f(\epsilon_{c})=1-F_0^2$, where
$F_n=(b z^n- b^{-1} z^{-n})/ (b - b^{-1})$
with $z=e^{-\beta h}$ and
$b=[(1+z e^{\beta \mu})/(1+z^{-1}e^{\beta \mu})]^{1/2}$.
Substitution of this result into the thermodynamic potential
(\ref{potential}) leads to:
$\Omega_{0}=-\beta^{-1}
\log\left( 1+ 2 e^{\beta \mu} \cosh(\beta h) \right)$.
We note that the impurity does
not contribute in this limit. $\Omega_0$ is just the thermodynamic potential
per unit length for a lattice where each site has three possible states
\cite{Schlottmann}.
In this case the entropy, specific heat, etc. are easy to interpret.
To higher order in $\beta$,
the quasi-energies
are no longer constants and we expand them as
$\beta \epsilon_{sj}(x)=\beta\epsilon_{sj}^0(x)+\sum_{n=1}^{\infty}
(\pi \beta J)^n \zeta_{sj}^{(n)}(x)$, and
$\beta \epsilon_{c}(x)=\beta\epsilon_{c}^0(x)+\sum_{n=1}^{\infty}
(\pi \beta J)^n \zeta_{c}^{(n)}(x)$,
where $\beta\epsilon_{sn}^0=\log(F_{n-1}F_{n+1})$ and
$\beta\epsilon_{c}^0=\log(F_0^2/(1-F_0^2))$.
Below we discuss the lightly doped Mott-Hubbard insulator.
Near half-filling, we have solved for
$\zeta_{sj}^{(1)}(x)$ and $\zeta_{c}^{(1)}(x)$ and their
Fourier transforms are
${\widehat\zeta}_{sn}^{(1)}(\omega)=
-(F_0/ F_1)\left[ (F_n/F_{n-1}) e^{-n|{\omega|\over 2}}
-(F_n/ F_{n+1}) e^{-(n+2){|\omega|\over 2}}\right]\;\;$, and
${\widehat\zeta}_{c}^{(1)}(\omega)= - (1/ F_1^2) e^{-|\omega|}\;\;$.
Substituing this into (\ref{potential}), leads
to the thermodynamic potential
$\Omega=\Omega_0+\Omega_1 +O((\beta J)^2)$, where
$\Omega_1=\Omega_1^{h}+\Omega_1^{i}/L=
-(J/F_1^2)\left(1-g/L\right)$.
{}From the latter we infer
immediately that
\begin{equation}
{\Delta S^i\over\Delta S^h}={\Delta C_v^i\over \Delta C_v^h}=
{\Delta M^i\over\Delta M^h}={\Delta\chi_s^i\over\Delta\chi_s^h}
={\Delta\chi_c^i\over\Delta\chi_c^h}=-g\;\;,
\end{equation}
where $\Delta S^{i,h}\;,\Delta C_v^{i,h}\;,\Delta M^{i,h}\;,
\Delta\chi_s^{i,h}\;,\Delta\chi_c^{i,h}\;$ denote the $O(\beta J)$
corrections to the impurity (host)
entropy, specific heat, magnetization, spin susceptibility, and
compressibility,
respectively. The simplicity of this formula is remarkable.
It expresses the fact that in the
weak coupling limit, the impurity ``mimics'' the host chain.
Notice that at $g=1$, the expression for the thermodynamic
potential density ( remember to multiply by $L/(L-1)$)
reduces to that of a chain of $L-1$ sites.
We have evaluated explicitly the entropy, specific heat, etc. to $O(\beta J)$
as a function of $\mu$, $h$ and $T$. For brevity, we mention
below only the simplest results\cite{Bares}.
Inverting the equation that defines the density $\rho=1-\delta$,
($\delta=$ hole density) as function of $\mu$, yields for $\delta\ll 1$
\begin{equation}
\mu=-{\beta}^{-1}\left[ \log \delta +\log (1+\delta) +
\log(2 \cosh(\beta h))+ \beta J \left(1-{g\over L}\right)\right]
+O(\delta^3,(\beta J)^2)\;\;.
\label{chemical}
\end{equation}
In the Sutherland representation,
the Heisenberg chain is obtained for
$\mu\rightarrow \infty$ ( remember that $\mu$ plays the role
of the chemical potential
for the holes).
Substituing (\ref{chemical})
back into the entropy per unit length, at $h=0$, we have
\begin{eqnarray}
{S\over k_B}
&\approx&\log 2 +\delta (1-[\log \delta +\log 2]) - {\delta^2\over
2}\nonumber\\
&+&{\beta J\over 2} (1-{g\over L})\{ \delta [\log\delta + \log 2 ]+
\delta^2 ( 1-2 [\log \delta +\log 2] ) \}+O(\delta^3,(\beta J)^2)\;\;,
\label{entropyzero}
\end{eqnarray}
The various contributions to the entropy are easy to interpret:
(i) the entropy of free spins $1/2$
(of density $1-\delta$);
(ii) the entropy of free holes; (iii)
a reduction of the entropy due to the exchange
correlations ($J>0$ and $\log\delta < -\log 2$ for $\delta\ll 1$);
(iv) an increase in entropy due to the impurity that depletes locally
the exchange.
The specific heat in zero field near half-filling behaves as
\begin{equation}
C_v = k_B\delta [\log\delta +\log 2]\left[[\log\delta +\log 2]
-\beta J\left(1-{g\over L}\right)\left(1+ {1\over 2}[\log\delta +\log 2]
\right)\right]
+O(\delta^3,(\beta J)^2)\;,
\end{equation}
The magnetization is finite in an external field and
near half-filling takes the form
\begin{equation}
{M\over\mu_B}=(1-\delta ) \tanh(\beta h)-\beta J \left( 1- {g\over L}\right)
(1-\tanh^2(\beta h))(1-3\delta+3\delta^2)+O(\delta^3,(\beta J)^2)\;,
\end{equation}
i.e., the magnetization is suppressed by (i) the doping that
reduces the number of spins; (ii) the antiferromagnetic correlations due
to the exchange. To next order, the defect
enhances the magnetization locally, while the doping
renormalizes down the effective exchange between nearest neighbors.
The $\delta^2$ term can be interpreted as an
attractive interaction between the holes that effectively increases the
renormalized exchange and thereby penalizes the magnetic state.
We have also evaluated the charge and spin susceptibilities to $O(\delta^2)$.
The general expressions are rather complex and will be discussed
elsewhere\cite{Bares}.

We now pass to the low temperature region, i.e., strong coupling regime.
We note that the quasi-energies $\epsilon_{sn}(x)>0$ for $n\geq 2$, $h>0$
(compare with ref. \cite{Takahashi}) and so
as $T\rightarrow 0$, only $\epsilon_{s1}(x)=\epsilon_{s}(x)$ and
$\epsilon_{c}(x)$ survive in equations (\ref{thermoa}) and (\ref{thermob})
\cite{Comment1}.
At $T=0$, the integral equations determining $\epsilon_{s}(x)$
and $\epsilon_{c}(x)$ coincide with those of ref. \cite{BaresCollabor}.
We have explicitly evaluated the distribution of roots as well as the
quasi-energies in zero field
near half-filling \cite{Bares}. The first non-trivial
corrections to the spinon and holon quasi-energies appear to third order
in the density of holes $\delta_0$ ($\delta=\delta_0+\delta_i/L$), i.e.,
\begin{mathletters}
\begin{eqnarray}
\epsilon_s(x)&\approx &-{\pi J\over 2 \cosh(\pi x)}
\left[1+\delta_0^3\left({\pi\over 2 \log 2}\right)^3
\left({\zeta(3)\over\pi}- {5\pi\log 2\over 16}[1-2 \tanh^2(\pi x)]
\right)\right]\;\;,
\label{energys}\\
P_s(x)&\approx & {\pi\over 2} -\tanh^{-1}(\sinh(\pi x))(1+\delta_0)
\nonumber\\
&+&
{\delta_0^3\over 3}\left({\pi\over 2\log 2}\right)^3
\left[ \pi\log 2 \;{\tanh(\pi x)\over \cosh(\pi x)}-
{3\zeta(3)\over 2\pi}+\left( {2\log 2\over \pi}\right)^3\right]\;\;,
\label{momentums}
\end{eqnarray}
\end{mathletters}%
\begin{mathletters}
\begin{eqnarray}
\epsilon_c(x)&\approx &\pi J \left[ {\log 2\over \pi}-R(x)+
\delta_0^2 \; {3\zeta(3)\over 4\pi}
+\delta_0^3 \left( {3\zeta(3)\over 2\pi}
\left({\pi\over 2 \log 2}\right)^2
-{\zeta(3)\over\pi} R(x)-{5\log 2\over 16\pi}R''(x)\right)\right]\;,
\label{energyc}\\
P_c(x)&\approx & 2\pi F(x)\left[1+\delta_0-\delta_0^3 \;{\zeta(3)\over 2\pi}
\left({\pi\over 2\log 2}\right)^3\right]
-\delta_0^3\;{\pi\over 3} \left({\pi\over 2\log 2}\right)^2 R'(x)\;,
\label{momentumc}
\end{eqnarray}
\end{mathletters}%
where $P_{c(s)}(x)$ denotes the momentum of the $c(s)$-type excitation relative
to the ground-state, $\zeta(3)$ is the Riemann zeta function,
$R(x)=\Re{\rm e}
\left[\Psi(1+ix/2)-\Psi([1+ix]/2)\right]/2\pi$ with the definition
$\Psi(x)=d\log\Gamma (x)/dx$.
$R'(x)$ and $R''(x)$ denotes the first and
second derivatives of $R(x)$ while
$2\pi F(x)=(\log\pi)/2+\log\left(\Gamma(1+ix/2)/\Gamma([1+ix]/2)\right)$.
The $O(\delta_0^2)$ term in $\epsilon_c(x)$ is due to a
chemical potential correction, i.e.,
$\mu=J[\log 2-  3\zeta(3)\delta_0^2/4]+O(\delta^4)$,
that shifts the bottom of the holon spectrum upwards.
{}From (\ref{energys}) and (\ref{energyc}), we see that the spinons
and holons behave as well defined
objects with respective renormalized velocities
$v_c\approx 3\pi \zeta(3) J\delta_0/ 8 \log^22$ and
$v_s\approx \pi J (1-\delta_0)/2$, to $O(\delta^2)$. Thus,
the holons and spinons
scatter off the impurity one by one. This
will be important when we deal with low-temperature transport properties
\cite{Bares}. In zero field,
the hole-charge is depleted in the region surrounding the defect, i.e.,
the impurity is screened
by an amount $\delta_i=-G(\theta) \delta_0$,
where $G(\theta)=1-\pi R(\theta)/log 2 >0$, while
the magnetization vanishes locally, i.e., the impurity is magnetically inert.
This redistribution of charge
around the defect explains why the spinon- and holon-impurity phase shifts
can simply be expressed in terms of the holon-holon and spinon-holon
phase shifts\cite{Bares}.

The zero-field ratios of impurity to host
susceptibilities can be shown
to be $\chi_{\alpha}^i/\chi_{\alpha}^h=
[\rho_{\alpha}^{i}(q_{\alpha}^0)+\rho_{\alpha}^{i}(-q_{\alpha}^0)]/
2\rho_{\alpha}^{h}(q_{\alpha}^0)$, i.e., equal to
the ratio of the average (over the left and right Fermi-points)
density of states
of the impurity
to the host chain. Near half-filling, we find $\chi_{s}^i/\chi_{s}^h
\approx -\delta_0 G(\theta)$, i.e., the local magnetic response in
zero field is controlled by the screening process,
while for the compressibility ratio we have
 $\chi_{c}^i/\chi_{c}^h
\approx G(\theta)+\delta_0^2 (\pi/2\log2)^2 [(\pi/2\log 2)R''(\theta)
+(3\pi \zeta(3)/2 \log^2 2) R(\theta)+2R'(\theta)]$.
At finite magnetic field, the defect becomes magnetically actif. For
weak fields, $h\ll J$, and near half-filling,
the Wiener-Hopf method leads to a local magnetization
that is linear in $h$, and
$M_i/M_h\approx
e^{\pi\theta}-1-\delta_0 [e^{\pi\theta}-(\pi R(\theta)/\log 2)]$, for
$\theta\ll|\log([e /2\pi] h)|/\pi$.
Notice that the impurity-bulk coupling is infrared finite in contrast to
the Kondo problem (compare with ref. \cite{TsvelickWiegmann}).
At zero temperature, we can derive simple relations between the $h=0$
susceptibility ratios $\chi_{\alpha}^i/\chi_{\alpha}^h $
and their finite field counterparts,
$\chi_{\alpha}^i(h)/\chi_{\alpha}^h(h) $, i.e.,
\begin{mathletters}
\begin{eqnarray}
{\chi_c^i(h)\over\chi_c^h(h)}&=&
\left[1+ {v_c\over v_s}{Z_{cs}^2\over Z_{cc}^2}\right]^{-1}
\left[{\chi_c^i\over\chi_c^h}+{\chi_s^i\over\chi_s^h}
{v_c\over v_s}{Z_{cs}^2\over Z_{cc}^2}\right]\;\;,
\label{suscepa}\\
{\chi_s^i(h)\over\chi_s^h(h)}&=&
\left[1+ {v_c\over v_s}
\left({ 2 Z_{ss}-Z_{cs}\over 2 Z_{sc}-Z_{cc}}\right)^2
\right]^{-1}
\left[{\chi_c^i\over\chi_c^h}+ {\chi_s^i\over\chi_s^h}{v_c\over v_s}
\left({ 2 Z_{ss}-Z_{cs}\over 2 Z_{sc}-Z_{cc}}\right)^2
\right]\;\;,
\label{suscepb}
\end{eqnarray}
\end{mathletters}%
where $v_{\alpha }$, $\alpha=c,s$, denotes the velocity of the $\alpha$
pseudoparticle at the $T=0$ Fermi points $\pm q_{\alpha}^0$ for finite $h$ and
$Z_{\alpha\beta}$ the ``dressed charge matrix''(essentially
the S-matrix of the s- and c- pseudoparticles at the Fermi surface,
see ref.\cite{BaresCollabor}). The formulas (\ref{suscepa}) and
(\ref{suscepb}) are universal in the sense that only the coefficients
in front of the matrix elements $Z_{\alpha\beta}$ depend
on the model under consideration \cite{Bares}.

We have evaluated the low-temperature corrections to the
thermodynamic potential
to $O(T^4)$, from which we have inferred the low-T magnetization, entropy,
susceptibilities, a.s.o. \cite{Bares}.
We quote the specific heat $C_v=C_v^{(1)}+C_v^{(2)}
+O(T^4)$, with
\begin{mathletters}
\begin{eqnarray}
C_v^{(1)}&=& {\pi k_B^2\over 3} T\sum_{\alpha}{1\over v_{\alpha }}
\left(1+{1\over L}
{{\rho_{\alpha i}(q_{\alpha}^0)+\rho_{\alpha i}(-q_{\alpha}^0)}
\over 2\rho_{\alpha h}(q_{\alpha}^0)}\right)\;\;,
\label{heat1}\\
C_v^{(2)}&=& {\pi^3 k_B^4\over 3} T^3 \sum_{\alpha}{1\over v_{\alpha }}
[ A_\alpha\left( 1+{1\over L}
{{\rho_{\alpha i}(q_{\alpha}^0)+\rho_{\alpha i}(-q_{\alpha}^0)}
\over 2\rho_{\alpha h}(q_{\alpha}^0)}\right)\nonumber\\
&+& B_{\alpha}{\rho_{\alpha h}'(q_{\alpha}^0)\over\rho_{\alpha
h}(q_{\alpha}^0)}
\left( 1+{1\over L}
{{\rho_{\alpha i}'(q_{\alpha}^0)+\rho_{\alpha i}'(-q_{\alpha}^0)}
\over 2\rho_{\alpha h}'(q_{\alpha}^0)}\right)\nonumber\\
&+& C_{\alpha} {\rho_{\alpha h}''(q_{\alpha}^0)
\over\rho_{\alpha h}(q_{\alpha}^0)}
\left( 1+{1\over L}
{{\rho_{\alpha i}''(q_{\alpha}^0)+\rho_{\alpha i}''(-q_{\alpha}^0)}
\over 2\rho_{\alpha h}''(q_{\alpha}^0)}\right)]\;\;.
\label{heat2}
\end{eqnarray}
\end{mathletters}%
The ratios of zeroth, first and second derivatives of the
Fermi surface density of states of the impurity to the
host chain occur in (\ref{heat1}) and (\ref{heat2}).
The coefficients
$A_{\alpha}$,$B_{\alpha}$ and $C_{\alpha}$ depend in a universal way
on the Fermi surface
velocity,
the curvature of the spectrum and its rate of change as well
as on the Scattering matrix and its higher derivatives.
For the sake of brevity, we omit here these lengthy and complicated expressions
\cite{Bares}.
The linear term in $T$, (\ref{heat1}), coincides with that inferred from the
zero temperature finite size spectrum calculations\cite{Bares}, and
can alternatively be derived
from boundary conformal field theory \cite{AffleckCollabor}.
The next order term (\ref{heat2}) cannot be calculated within the framework
of conformal field theory.
Very close to half-filling, equations
(\ref{heat1}) and (\ref{heat2}) do not hold, however.
At low $T$, when the charge carrier density $\delta_0$ is small, a crossover
takes place to a regime where the holon-liquid behaves as
a non-degenerate 1D Fermi system coupled to an impurity.
A crude estimate for the onset of the crossover is
obtained when the width of the holon ``Fermi-sea'' is of order $k_B T$, i.e.,
$\delta_c\approx [k_B T/J]^{1/2} (4 \pi^{-1}\log 2 \;(3 \zeta(3))^{-1/2})$.
For instance, at $T=10$K with
an exchange of order $1$ eV, $\delta_c\approx 10^{-2}$ while
at room temperature we have $\delta_c\approx 10^{-1}$ \cite{Bares}.
A similar crossover takes place for the spinon liquid near the
transition to the ferromagnetic state.
Following ref.\cite{FendleyLS,Tsvelick}
we can calculate the conductance through the impurity
at low-T. We shall report on this elsewhere\cite{Bares}.

In summary, we have proposed a novel strategy
to construct solvable models with
impurities. The hamiltonians generated by our method contain terms
that are difficult to justify on physical grounds. Yet some of those terms
are, in the models considered so far, and
should generically be, irrelevant to the low-energy properties of the model.
We have illustrated the idea by performing explicit calculations
in a simple model with a single impurity. We have derived various results
that are beyond
boundary conformal field theory by solving the thermodynamic Bethe ansatz
equations in the high-and low-temperature limits.
The transport properties of these class of models will be explored
in future work.


It is a pleasure to thank P. Nozi\`eres and A. Gogolin for discussions and
encouragements.
We wish to thank the organizers
of a workshop on exact solutions at Aspen for the inspiring
atmosphere that stimulated this work. We also gratefully acknowledge
the Aspen Center for Physics for the kind hospitality
during a stay in July 1994.


\begin{references}
\bibitem{Comment1} To simplify the discussion, we choose a finite
magnetic field. At $h=0$, we have an additional branch of excitations
parametrized by complex roots (two-strings) that is degenerate
with $\epsilon_s(x)$. A complete and careful
discussion of this point can be found in
ref. \cite{BaresCollabor}.
\end{references}
\end{document}